\documentclass[a4paper]{aa}

\usepackage{graphicx,natbib}
\usepackage{txfonts}
\usepackage{subfigure}
\usepackage{dblfloatfix}
\usepackage{fixltx2e}
\usepackage{color}

\usepackage{arydshln}

\newcommand{\kms}{km\,s$^{-1}$}

\newcommand{\figps}[1]{\resizebox{\hsize}{!}{\rotatebox{0}{\includegraphics{#1}}}}

\newcommand{\filps}[1]{\resizebox{\hsize}{!}{\rotatebox{90}{\includegraphics{#1}}}}

\begin{document}

\title{Magnetic field topology of the cool, active, short-period \\ binary system $\sigma^2$~Coronae Borealis%
\thanks{Based on observations obtained at the Bernard Lyot Telescope (TBL; Pic du Midi, France) of the Midi-Pyr\'{e}n\'{e}es Observatory, which is operated by the Institut National des Sciences de l'Univers of the Centre National de la Recherche Scientifique of France. Also based on observations obtained at the Canada-France-Hawaii Telescope (CFHT), which is operated by the National Research Council of Canada, the Institut National des Sciences de l'Univers of the Centre National de la Recherche Scientifique of France, and the University of Hawaii.}}

\author{L.~Ros\'{e}n  \inst{1}
   \and O.~Kochukhov \inst{1}
   \and E.~Alecian \inst{2}
   \and C.~Neiner \inst{3}
   \and J.~Morin \inst{4}
   \and G.A.~Wade \inst{5}
   \and the BinaMIcS collaboration
  }
\institute{Department Physics and Astronomy, Uppsala University, Box 516, 751 20 Uppsala, Sweden
        \and Universit\'e Grenoble Alpes, CNRS, IPAG, F-38000 Grenoble, France
        \and  LESIA, Observatoire de Paris, PSL Research University, CNRS, Sorbonne Universit\'{e}s,
                UPMC Univ. Paris 06, Univ. Paris Diderot, Sorbonne Paris Cit\'{e}, 5 place Jules Janssen, 92195
                Meudon, France
        \and LUPM, Universit\'{e} de Montpellier, CNRS, Place Eug\`{e}ne Bataillon, 34095, France
        \and Department of Physics and Space Science, Royal Military College of Canada, P.O. Box 17000, Station `Forces', Kingston, Ontario, Canada, K7K 7B4
}

\date{Received 3 August 2017 / Accepted 29 January 2018}

\titlerunning{Magnetic field topology of $\sigma^2$~CrB}

\abstract
{}
{The goal of this work is to study the cool, active binary star $\sigma^2$~CrB, focussing on its magnetic field. The two F9--G0 components of this system are tidally locked and in a close orbit, increasing the chance of interaction between their magnetospheres.}
{We used Stokes $IV$ data from the twin spectropolarimeters Narval at the TBL and ESPaDOnS at the CFHT. The least-squares deconvolution multi-line technique was used to increase the signal-to-noise ratio of the data. We then applied a new binary Zeeman-Doppler imaging code to reconstruct simultaneously the magnetic topology and brightness distribution of both components of $\sigma^2$~CrB. This analysis was carried out for two observational epochs in 2014 and 2017.}
{A previously unconfirmed magnetic field of the primary star has been securely detected. At the same time, the polarisation signatures of the secondary appear to have a systematically larger amplitude than that of the primary. This corresponds to a stronger magnetic field, for which the magnetic energy of the secondary exceeds that of the primary by a factor of 3.3--5.7. While the magnetic energy is similar for the secondary star in the two epochs, the magnetic energy is about twice as high in 2017 for the primary. The magnetic field topology of the two stars in the earlier epoch (2014) is very different. The fraction of energy in the dipole and quadrupole components of the secondary are similar and thereafter decrease with increasing harmonic angular degree $\ell$. At the same time, for the primary the fraction of energy in the dipole component is low and the maximum energy contribution comes from $\ell=4$. However, in the 2017 epoch both stars have similar field topologies and a systematically decreasing energy with increasing $\ell$. In the earlier epoch, the magnetic field at the visible pole appears to be of opposite polarity for the primary and secondary, suggesting linked magnetospheres. The apparent rotational periods of both $\sigma^2$~CrB components are longer than the orbital period, which we interpret as an evidence of a solar-like differential rotation.} 
{Despite their nearly identical fundamental parameters, the components of $\sigma^2$~CrB system exhibit different magnetic field properties. This indicates that the magnetic dynamo process is a very sensitive function of stellar parameters.}
\keywords{polarisation -- stars: activity -- stars: magnetic fields -- stars: late-type -- stars: individual ($\sigma^2$~CrB)}

\maketitle

\section{Introduction}
\label{intro}

Magnetic fields are omnipresent in the universe and are therefore involved in many astrophysical systems. However, the impacts of a magnetic field are poorly understood and are often not even taken into account in the modelling of many processes. Specifically for stars, magnetic fields are known to play a key role at all stages of stellar evolution since they, for instance, affect accretion, slow stellar rotation, influence mass loss, and cause  increased emission of high energy radiation and particles. These phenomena also affect the environment surrounding the star and can have a significant impact on planetary atmospheres and other nearby stars. 
 
To deepen the understanding of how magnetic fields influence a star, and vice versa, it is important to disentangle the magnetic effects from other processes. One way to accomplish this is to study binary stars, for which the components can be assumed to have been formed simultaneously from the same molecular cloud, giving the stars the same initial conditions and age. An international collaboration Binarity and Magnetic Interactions in various classes of Stars \citep[BinaMIcS,][]{Alecian2015} aims to investigate binary systems containing magnetic stars in order to study the relation between binarity and magnetism in stars at various stages of evolution. 

More specifically, the BinaMIcS collaboration focusses on a few key questions. Binary stars affect each other, especially if they are close. Tidal forces acting on the two stars impacts their internal flows \citep[e.g.][]{Ogilvie2007,Remus2012}. This, in turn, can influence fossil fields or perhaps drive a magnetic dynamo \citep{Barker2014,Cebron2014}. If both components of the binary are magnetic, their magnetospheres might interact or perhaps even have connected magnetic field lines forming a joint magnetosphere \citep[e.g.][]{Holzwarth2015}. In addition, a stellar magnetic field influences the stellar wind, and hence the angular momentum loss. In a binary system, this effect might be further increased or altered. 

In order to properly investigate these effects, a large sample of stars is needed. For cool stars, the selection of BinaMIcS targets was based on previous studies where signs of magnetic activity have been found. A sample of about 20 such systems was selected for more comprehensive observations. A requirement was set that the binary components are close, i.e. the orbital period is $\lesssim 20$~d. 

The first step in a detailed study of the magnetic activity effects is mapping the stellar surface. So far, starspots have only been mapped on a few stars other than the Sun through direct imaging using interferometry \citep{Parks2015,Roettenbacher2016}. However, interferometry is limited to bright and spatially extended targets, which is usually not the case for most stars. Also, direct imaging provides a map of brightness or temperature on the stellar surface, which, although plausibly connected to a magnetic field, does not provide any direct information about the field strength and topology. Therefore, indirect imaging techniques are used to investigate both the temperature/brightness and magnetic field at the surface of a star. One such commonly used method is Zeeman-Doppler imaging \citep[ZDI;][]{Brown91,Kochukhov16}. This technique uses the rotational Doppler broadening of spectral line intensity and polarisation profiles, where one side of the profile is red-shifted and the other side is blue-shifted. As a consequence of this broadening, each point in the spectral line profile corresponds to a vertical stripe on the stellar surface. 
Cool spots therefore cause a distortion in the disc-integrated intensity profile, while magnetic spots give rise to a signature in the polarisation spectrum. Using a time series of spectropolarimetric observations, it is possible to identify and trace the spots as the star rotates, and ultimately reconstruct detailed 2D temperature and magnetic field maps.

In this study, we have obtained and analysed a spectropolarimetric data set of the cool, active, short-period, double-line binary star $\sigma^2$~CrB (TZ~CrB, HR\,6063, HD\,146361). The orbit of this system has been resolved interferometrically, providing, together with the spectroscopic orbit, useful constraints on the fundamental stellar parameters and orbital inclination \citep{Raghavan2009}. Both components have effective temperatures slightly hotter than the Sun, that is, $6000\pm50$~K and $5900\pm50$~K for the primary and secondary, respectively \citep{Strassmeier2003}. Both components are rapid rotators and have in previous studies been shown to exhibit cool surface spots \citep{Strassmeier2003}. Direct evidence of a magnetic field on the secondary component has been obtained through a Stokes $V$ polarisation profile analysis \citep{Donati1992}, however no quantitative information about the field strength or topology could be extracted from these early polarisation observations. In addition, the two components are close together and have an orbital period of only about 1.14 days, making this system very interesting for studies of magnetospheric and tidal interaction.

Our paper is structured as follows. We describe the observational data, telescopes, and instrumentation used in Sect.~\ref{obs}. In Sect.~\ref{multi} we discuss magnetic field detection and calculation of the least-squares deconvolved (LSD) profiles. In Sect.~\ref{param} we present derivation of improved stellar and orbital parameters. In Sect.~\ref{zdi} we describe a new ZDI code that we have used to reconstruct the surface brightness and magnetic field distributions for both components of $\sigma^2$~CrB. The resulting maps are presented in Sect.~\ref{res}, and the results are discussed in Sect.~\ref{dis}.  

\begin{table*}
\caption{Log of Narval (2013--2014) and ESPaDOnS (2017) observations of $\sigma^2$~CrB.}
\label{obsd}
\centering
\begin{tabular}{lccrrc}
\hline\hline
Date & HJD & S/N$_{\rm peak}$   & RV$_{\rm p}$~~ & RV$_{\rm s}$~~ & Phase  \\
(UTC) & (2,400,000 +) & (pixel$^{-1}$)   & (\kms) & (\kms) &  ($T_{\rm conj}^{\rm s \rightarrow p}+P_{\rm orb}E$) \\
\hline
2013 May 12 &   56425.36651  & 820&  -50.15 &  25.42 &    0.104  \\
2013 May 12 &   56425.53576  & 845&  -74.32 &  50.58 &    0.253  \\
2013 Jun 15 &   56459.40058  & 656&    0.62 & -26.61 &   29.964  \\
2014 May 07 &   56785.48672  & 603&  -33.95 &   9.22 &  316.057  \\
2014 May 07 &   56785.54913  & 340&  -51.80 &  27.84 &  316.112  \\
2014 May 07 &   56785.65323  & 479&  -70.78 &  47.25 &  316.203  \\
2014 May 08 &   56786.46289  & 608&   19.39 & -44.79 &  316.914  \\
2014 May 08 &   56786.57382  & 739&  -16.70 &  -8.49 &  317.011  \\
2014 May 08 &   56786.63995  & 742&  -38.26 &  13.36 &  317.069  \\
2014 May 09 &   56787.45919  & 647&   47.34 & -73.32 &  317.788  \\
2014 May 09 &   56787.56082  & 731&   30.69 & -56.34 &  317.877  \\
2014 May 09 &   56787.62626  & 734&   12.28 & -37.84 &  317.934  \\
2014 May 14 &   56792.40547  & 685&  -56.40 &  32.19 &  322.128  \\
2014 May 14 &   56792.51830  & 699&  -73.06 &  49.39 &  322.227  \\
2014 May 14 &   56792.59004  & 737&  -72.07 &  48.14 &  322.289  \\
2014 May 15 &   56793.45615  & 682&  -31.16 &   5.91 &  323.049  \\
2014 May 15 &   56793.55307  & 746&  -58.26 &  34.05 &  323.134  \\
2014 May 15 &   56793.61798  & 708&  -69.60 &  45.76 &  323.191  \\
2014 May 16 &   56794.44004  & 655&   19.56 & -45.07 &  323.913  \\
2014 May 16 &   56794.55777  & 742&  -18.46 &  -6.73 &  324.016  \\
2014 May 16 &   56794.62533  & 750&  -40.26 &  15.54 &  324.075  \\
2014 May 17 &   56795.45261  & 718&   45.70 & -71.99 &  324.801  \\
2014 May 17 &   56795.57364  & 757&   21.49 & -47.16 &  324.907  \\
2014 May 17 &   56795.63518  & 742&    2.24 & -27.63 &  324.961  \\
2014 May 18 &   56796.43721  & 690&   40.23 & -66.56 &  325.665  \\
2014 May 18 &   56796.55238  & 611&   48.70 & -74.85 &  325.766  \\
2014 May 18 &   56796.61742  & 730&   42.58 & -68.80 &  325.823  \\
2014 Jun 02 &   56811.52323  & 667&   23.82 & -49.54 &  338.901  \\
2014 Jun 04 &   56813.44704  & 604&   19.74 & -45.84 &  340.588  \\
2014 Jun 07 &   56816.45051  & 552&  -72.88 &  50.41 &  343.224  \\
2014 Jun 10 &   56819.46148  & 500&   33.81 & -59.71 &  345.865  \\
2014 Jun 18 &   56827.50045  & 511&   18.50 & -44.11 &  352.918  \\
2014 Jun 19 &   56828.43844  & 605&   48.79 & -76.20 &  353.741  \\
2017 Jan 08 &   57762.16934  & 923&    3.99 & -30.11 & 1172.954  \\
2017 Jan 10 &   57764.15187  & 915&   45.29 & -72.04 & 1174.693  \\
2017 Jan 12 &   57766.16472  & 814&  -26.99 &   3.09 & 1176.459  \\
2017 Jan 13 &   57767.17351  & 981&  -63.23 &  39.94 & 1177.344  \\
2017 Jan 14 &   57768.16620  & 876&  -72.52 &  49.55 & 1178.215  \\
2017 Jan 15 &   57769.16826  & 906&  -47.13 &  23.49 & 1179.095  \\
2017 Jan 16 &   57770.16645  & 887&   -1.95 & -23.82 & 1179.970  \\
2017 Jan 19 &   57773.08965  & 591&    1.42 & -26.79 & 1182.535  \\
2017 Jan 20 &   57774.11569  & 935&  -36.39 &  12.22 & 1183.435  \\
2017 Jan 22 &   57776.14231  & 852&  -72.43 &  49.46 & 1185.213  \\
\hline
\end{tabular}
\tablefoot{Observational UTC date can be found in the first column, and the heliocentric Julian date in the second column. The third column lists the peak signal-to-noise ratio per CCD pixel. The fourth and fifth columns show the derived radial velocities of the primary and secondary star, respectively. The last column gives the orbital phase calculated using an epoch of conjunction. 
}
\end{table*}

\section{Observations}
\label{obs}
The BinaMIcS collaboration has had two large observational programmes running from early 2013 to early 2017 \citep{Alecian2015}. The two programmes included about 170~h and 600~h of observational time with the twin spectropolarimeters Narval at the 2~m T\'{e}lescope Bernard Lyot at Pic du Midi Observatory in France and ESPaDOnS at the 3.6~m Canada-France-Hawaii Telescope, respectively. 

$\sigma^2$~CrB was observed with both Narval and ESPaDOnS. These instruments are cross-dispersed \'{e}chelle spectrographs \citep{Donati03,Petit2008}. They operate in the optical spectrum covering essentially all wavelengths between 3700 and 10000 \AA\ with a resolving power of about 65000. They can be used in both a non-polarimetric and a polarimetric mode. The spectrographs are equipped with dual fibres, allowing the two orthogonal states of the polarised spectra to be recorded at the same time. All four Stokes parameters ($IQUV$) can be obtained using Narval and ESPaDOnS. 

Only Stokes $IV$ data of $\sigma^2$~CrB were available for this study. Three Narval observations were made for this star in 2013 between 12 May and 15 June, and another 30 Narval observations were obtained a year later between 7 May and 19 June 2014, adding up to 33 observations in total. Another 10 observations were obtained with ESPaDOnS between 8 January and 22 January 2017. Detailed information about the observation dates and the signal-to-noise (S/N) ratio of individual spectra can be found in Table~\ref{obsd}. 

All observations were automatically reduced by the Libre-ESpRIT software \citep{Donati97}. We then performed a continuum normalisation by fitting a global smooth function to the Stokes $I$ spectra using a set of IDL routines. Detailed description of this continuum normalisation procedure is presented in the Appendix~\ref{contin}.

\section{Magnetic field detections}
\label{multi}

Evidence of a magnetic field can be found in the intensity spectrum or in the polarisation spectra of a star. The presence of a magnetic field causes a splitting of the energy levels of the atom, leading to the corresponding splitting or broadening of the line profile in the unpolarised intensity spectrum. The stronger the field, the larger is the gap between energy levels. A strong field is therefore easier to detect since the wavelength separation of Zeeman components is larger. A related phenomenon is Zeeman polarisation of the light. For a non-magnetic star, the circular and linear polarisation spectra inside spectral lines would normally be flat, showing no signal. However, if a magnetic field is present, magnetically sensitive lines exhibit a characteristic signature in the polarisation spectra. 

In this study, we used circular polarisation (Stokes $V$) spectra to investigate the magnetic fields. Linear Zeeman polarisation (Stokes $QU$) is usually more difficult to detect since these signatures are about ten times weaker \citep{Kochukhov11,Rosen13}. However, despite a relatively high S/N of Narval and ESPaDOnS Stokes $V$ spectra of $\sigma^2$~CrB, it was not possible to see any clear circular polarisation signatures in individual lines. In order to significantly increase the S/N ratio of the observations, we applied the LSD multi-line technique \citep[][]{Donati97}. In this study, we used the LSD code developed by \citet{Kochukhov2010}. In total, 2516 lines with an intrinsic depth larger than 20\% of the continuum were used. Lines that are broader than the average line were also excluded, for example the hydrogen lines and the Na D doublet. We also masked out wavelength regions contaminated by the telluric absorption. The LSD line mask was derived using a {\sc marcs} stellar model atmosphere \citep{Gustafsson2008} with an effective temperature of 6000~K, $\log g=4.5$, microturbulent velocity of 2 \kms, and solar metallicity, together with the atomic line data extracted from {\sc vald3} \citep{Piskunov1995,Kupka1999,Ryabchikova15}. The LSD profiles were calculated with a velocity step of 1.8 \kms\ and were scaled for a mean wavelength of 5192 \AA\ and a mean Land\'{e} factor of 1.219. 

The Stokes $V$ profiles of a few individual lines in the composite spectrum of this binary system were previously investigated by \citet{Donati1992}. That study reported a secure detection of the magnetic field in the secondary component of $\sigma^2$~CrB. A definite detection is commonly defined as the false alarm probability \citep[FAP;][]{Donati1992} being smaller than 10$^{-5}$. We have achieved definite detections of polarisation signatures in LSD profiles for both primary and secondary components in all observations.

\section{Stellar and orbital parameters}
\label{param}

The $\sigma$ Coronae Borealis system is thought to contain five stars \citep{Raghavan2009} and possibly even more. In addition to the two close components of $\sigma^2$~CrB, there is another solar-like star, $\sigma^1$~CrB that has an estimated effective temperature of 5821~K, a mass of 0.770~$M_{\odot}$, and metallicity [M/H]\,=\,$-0.05$ \citep{Valenti2005}. The orbit of $\sigma^1$~CrB is relatively wide and has an estimated period of over 700 years according to \citet{Raghavan2009}. The latter authors also argue that another binary system, consisting of two M-dwarf stars, seems to be physically associated with the $\sigma$ Coronae Borealis system even though the separation is believed to be more than 14000 AU.

The two stars of $\sigma^2$~CrB are thought to be relatively young. \citet{Strassmeier2003} argued for an age of a few times $10^7$ years based on the comparison with theoretical evolutionary tracks and the high lithium abundance. \citet{Raghavan2009} estimated an age of 0.5--1.5~Gyr, where 0.1--3~Gyr is within a 1$\sigma$ error, using isochrones.

Previous studies of $\sigma^2$~CrB have shown that the two components are very similar. The primary appears to be more luminous and slightly hotter, with a temperature of about 6000~K, compared to the secondary, which has a temperature of about 5900~K \citep{Strassmeier2003,Raghavan2009}. \citet{Raghavan2009} estimated a radius of 1.244$\pm$0.050 $R_{\odot}$ for both components while \citet{Strassmeier2003} determined lower and somewhat different radii for the primary, 1.14$\pm$0.04 $R_{\odot}$, and secondary, 1.10$\pm$0.04 $R_{\odot}$, respectively. 

We derived a new orbital solution using radial velocities determined from our 43 high-quality spectra. Additionally, we also re-analysed the 46 radial velocity measurements from \citet{Raghavan2009}. To derive the radial velocities, we applied the procedure of spectral disentangling to the Stokes $I$ LSD profiles. The disentangling code, described by \citet{folsom:2010,folsom:2013a}, was adapted to treat LSD profiles instead of individual lines. This code obtains mean profiles of each binary component and the corresponding radial velocities assuming that spectral variability is due to the binary motion alone. The code starts with a guess of radial velocities for each orbital phase, for which we adopted the values predicted by the orbital solution published by \citet{Raghavan2009}, and then derives mean profiles by fitting spectra at all available phases simultaneously. Then the individual radial velocities are refined and the mean profiles are derived again. Several such iterations are carried out until convergence criteria are satisfied. The radial velocity measurements obtained with this analysis are given in Table~\ref{obsd}. 

\begin{figure}
\centering
\filps{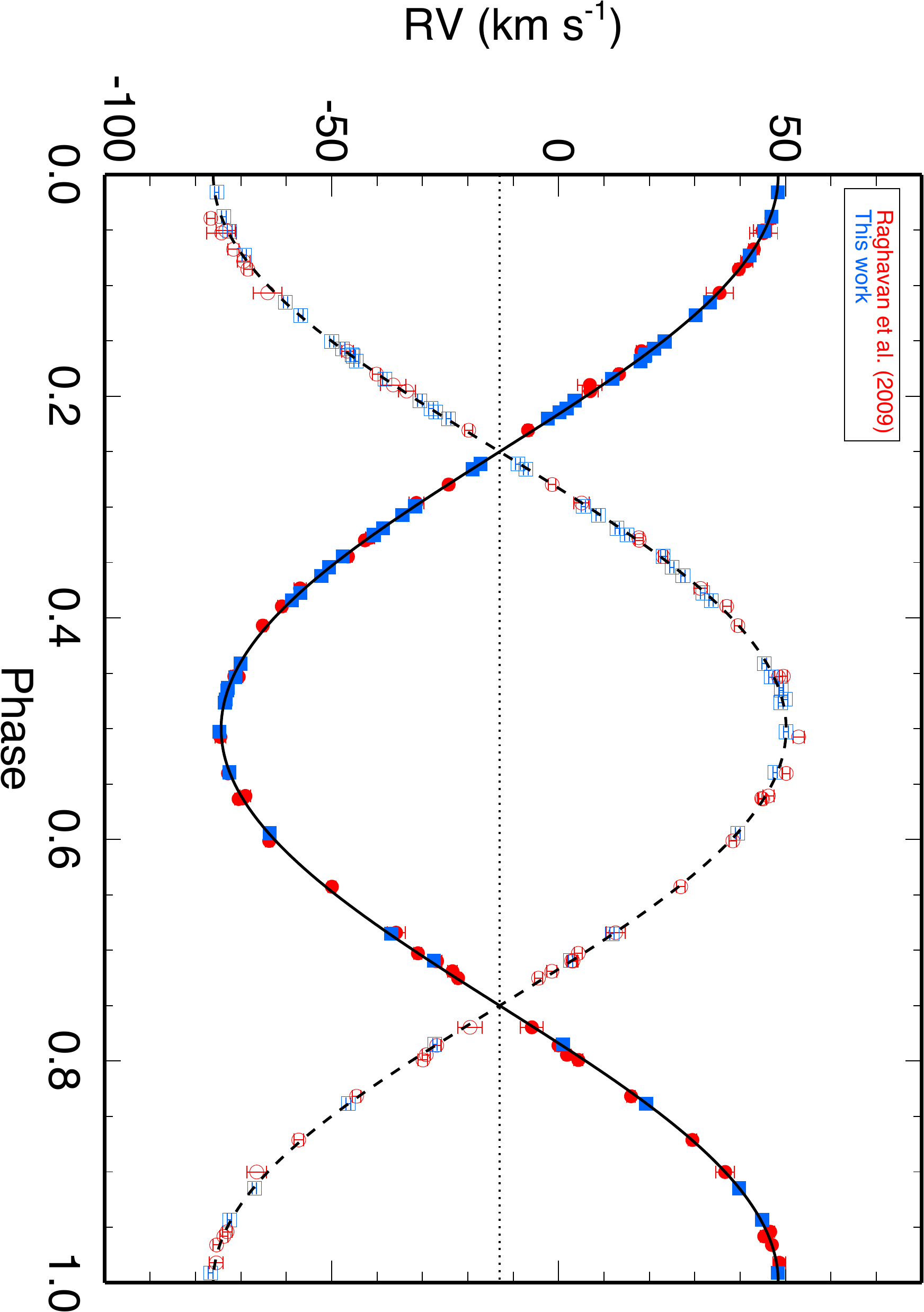}
\caption{Radial velocity of $\sigma^2$~CrB as a function of the orbital phase, calculated using $T_{\rm max}^{\rm p}$. The least-squares fit for the primary is shown with the solid line and for the secondary with the dashed line. The red circles are observations by \citet{Raghavan2009} and the blue squares are our new radial velocity measurements. The filled symbols represent the primary and the open symbols correspond to the secondary.}
\label{orbit_fit}
\end{figure}  

The orbital solution was then derived by performing a non-linear least-squares fit using the orbital parameters derived by \citet{Raghavan2009} as initial values and assuming a circular orbit. The resulting fits are presented in Fig.~\ref{orbit_fit}. The derived orbital period is 1.13979045$\pm$0.00000008 d, which is similar to the value given by \citet{Raghavan2009}. Adopting the orbital inclination of $28.08\degr\pm0.34\degr$ \citep{Raghavan2009}, the masses of the two components are found to be 1.107$\pm$0.013 $M_{\odot}$ and 1.077$\pm$0.012 $M_{\odot}$ for the primary and secondary, respectively. These values are in close agreement with the estimate by \citet{Strassmeier2003} (1.108$\pm$0.004 $M_{\odot}$ and 1.080$\pm$0.004 $M_{\odot}$). The derived orbital parameters are listed in Table~\ref{oparam}. The corresponding rms values of the orbital fit including all observations of the primary and secondary component are 0.83 and 0.88~\kms, respectively. In comparison, the rms of the fit from the \citet{Raghavan2009} study was 1.04 and 1.10~\kms\ for the primary and secondary. When only our radial velocity measurements are considered, the rms values are found to be 0.37 and 0.45~\kms, respectively. These low deviations from the fitted orbit demonstrate that our radial velocity measurements are not noticeably affected by the line profile distortions caused by cool spots. The orbital phases of all our observations, calculated relative to the conjunction epoch (phase 0.25 in Fig.~\ref{orbit_fit}), can be found in the last column of Table~\ref{obsd}. 

\begin{table}
\caption{Orbital parameters of $\sigma^2$~CrB derived in this study.}
\label{oparam}
\centering
\begin{tabular}{cc}
\hline\hline
Parameter & Value \\
\hline
$P_{\rm orb}$  (d) & 1.13979045 $\pm$ 0.00000008 \\
$T_{\rm max}^{\rm p}$  (HJD) & 2450127.6204  $\pm$ 0.0004  \\
$T_{\rm conj}^{\rm s \rightarrow p}$  (HJD) & 2450127.9054 $\pm$ 0.0004 \\
$K_{\rm p}$  (\kms)  & 61.366   $\pm$   0.097  \\
$K_{\rm s}$ (\kms)  & 63.106    $\pm$  0.098  \\
$\gamma$ (\kms)  & $-12.983$    $\pm$  0.106  \\
$a_{\rm p}\sin i $ $(R_\odot)$ & 1.383 $\pm$ 0.002  \\
$a_{\rm s}\sin i $ $(R_\odot)$ & 1.422  $\pm$ 0.002  \\
\hline
\end{tabular}
\tablefoot{Orbit is assumed to be circular. The parameter $T^{\rm p}_{\rm max}$ corresponds to the epoch of maximum primary velocity and $T_{\rm conj}^{\rm s \rightarrow p}$ corresponds to the conjunction epoch when the secondary is in front of the primary.}
\end{table}

\section{Zeeman-Doppler imaging}
\label{zdi}
\subsection{Method}

For the reconstruction of magnetic field topologies and brightness distributions on both components of $\sigma^2$~CrB, we used a new binary ZDI code {\sc InversLSDB}. This code is based on the {\sc InversLSD} code described by \citet{Kochukhov2014}, but extends the magnetic imaging problem to a composite spectrum containing contributions of two, possibly eclipsing, stars. 

{\sc InversLSDB} can model the surface structure of spectroscopic binary stars using one of the following two approaches. In the first case, the binary components are assumed to be tidally locked and co-rotating, and their surface shapes are described by Roche equipotentials. The calculation of corresponding non-spherical stellar surface grids, Doppler shifts, and visibilities of individual surface elements is performed with a set of routines (Piskunov \& Holmgren, private communication) following the treatment of the Roche-lobe geometry problem by \citet{Mochnacki1972}. In this case, the free parameters of the model include the orbital period, inclination angle, masses of the two components, and values of the corresponding Roche equipotentials or, equivalently, radii. This set of parameters allows one to establish Doppler shifts at each rotational phase without the need to specify the projected rotational velocities, $v \sin i$. In other words, the spectral line width in this model is controlled by the stellar radii. 
{\sc InversLSDB} also has another mode, in which the stars are treated as spherical bodies orbiting according to the prescribed, arbitrary eccentric orbit. The binary components can have individual rotational axis inclinations, along with different rotational periods and radii. Individual differential rotation of the components is implemented following previous surface mapping studies of single cool active stars \citep[e.g.][]{Donati97d,Petit02}. Both binary geometry modes of {\sc InversLSDB} were used for the present analysis of $\sigma^2$~CrB as described below.

The local Stokes parameter calculation used in our study assumed that LSD profiles behave as a normal spectral line with average parameters \citep[e.g.][]{BoroSaikia2015,Folsom2016}. The central wavelength and effective Land\'{e} factor were set equal to the  mean values of the LSD line mask applied to the observations (5193~\AA\ and 1.219). Furthermore, the line was assumed to split as a Zeeman triplet and its equivalent width was adjusted to fit the Stokes $I$ LSD profiles. The Milne-Eddington analytical solution of the polarised radiative transfer equation was used to calculate the synthetic local Stokes profiles. This single-line approach has been shown to work well for Stokes $IV$ spectra as long as the magnetic field strength is lower than a few kG \citep{Kochukhov2010}, which applies to most cool stars and to $\sigma^2$~CrB components in particular.

\begin{figure}
\centering
\figps{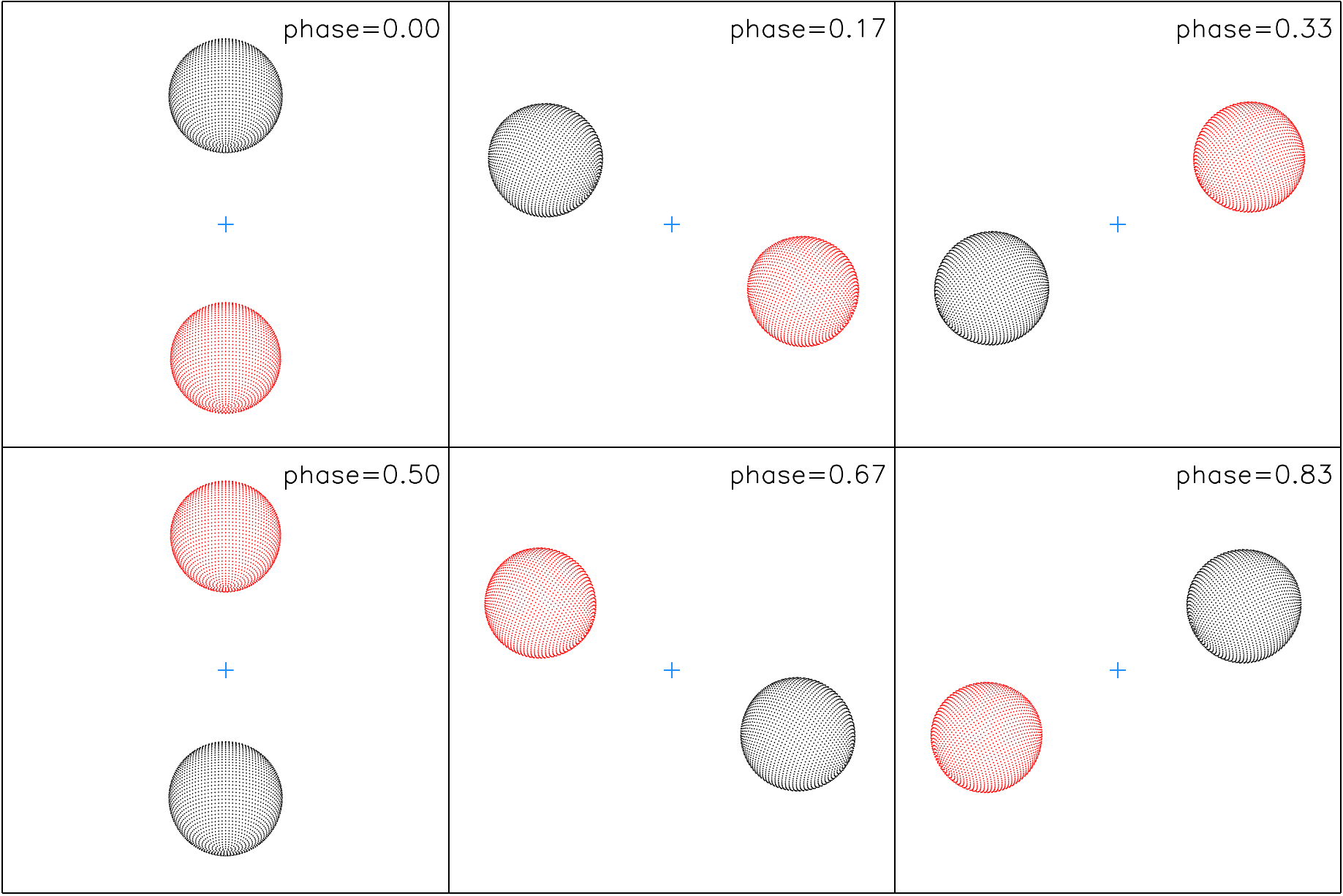}
\caption{Three-dimensional geometry of the $\sigma^2$~CrB system. The binary components are shown to scale relative to the size of the orbit. The six panels correspond to various orbital phases as indicated in each plot. The orbital inclination is $28\degr$. The shapes of the two stars are drawn according to the best-fitting Roche equipotential geometry. The surface of the primary  is shown with dark (black) points; the surface of the secondary is shown with light (red) points. The plus sign corresponds to the centre of mass position. }
\label{geo}
\end{figure} 

In this study we adopted the inclination angle of 28.08$\degr$ from \citet{Raghavan2009} together with our refined orbital period of 1.13979045~d and the component masses of 1.107~$M_{\odot}$ for the primary and 1.077~$M_{\odot}$ for the secondary, respectively. We then carried out the analysis of the Stokes $I$ LSD profiles using {\sc InversLSDB} in the Roche-lobe geometry mode. This modelling allowed us to establish the equivalent volume radii, 1.136~$R_\odot$ and 1.104~$R_\odot$ for the primary and secondary, respectively, best matching the Stokes $I$ LSD spectra. In addition, it was necessary to scale the primary spectra by a factor of 1.25 to account for the relative line strength difference between the components.

The successful application of the Roche-lobe geometry model to the Stokes $I$ data rules out significant misalignment of the orbital and rotational axes. The radii inferred by our analysis agree within error bars with the values obtained by \citet{Strassmeier2003}. An illustration of the derived 3D geometry of the system is presented in Fig.~\ref{geo}. In agreement with \citet{Strassmeier2003}, we found an insignificant ($\le$\,1.1\%) deviation from the spherical shapes. 

The surface magnetic field structure of each star was described by a superposition of poloidal and toroidal components, expressed in terms of spherical harmonic functions (without accounting for non-sphericity of the stars). The radial and horizontal poloidal components were represented by the two sets of harmonic expansion coefficients, $\alpha_{\ell,m}$ and $\beta_{\ell,m}$, and the horizontal toroidal component was defined by the third set, $\gamma_{\ell,m}$ \citep[see][]{Kochukhov2014}. The subscripts $\ell$ and $m$ correspond to the angular degree and the azimuthal order of each mode, respectively. 

In order to find a unique solution of an inverse problem using the Stokes $IV$ data, a regularisation function is required. High-order modes, corresponding to a complex magnetic field, were suppressed using a penalty function
\begin{equation}
R_B=\Lambda_B \sum_{\ell=1}^{\ell_{\rm max}}\sum_{m=-\ell}^{\ell} \ell^2(\alpha_{\ell,m}^2+\beta_{\ell,m}^2+\gamma_{\ell,m}^2),
\end{equation}
where $\Lambda_B$ is the magnetic regularisation parameter. The maximum angular degree was set to $\ell_{\rm max}=12$. In this study we did not try to adjust $\ell_{\rm max}$ to restrict the solution; instead it was chosen so that modes with $\ell \ge \ell_{\rm max}$ contain negligible magnetic energy. Given the moderately large $v \sin i$ of the targets (23--24~\kms) and the spectral resolution of the data ($FWHM$\,=\,4.6~\kms), we are potentially sensitive to modes with $\ell$ up to $4v\sin i / FWHM\approx 20$.

In addition, regularisation should also be applied to obtain a unique brightness distribution. Assuming that there should be no sharp contrasts in brightness between neighbouring surface zones, we implemented the Tikhonov regularisation 
\begin{equation}
R^{(1)}_T = \Lambda^{(1)}_T\sum_i \sum_j (T_i - T_{j(i)})^2
\end{equation}
in order to dampen such differences. In this equation index $i$ runs over all surface elements while index $j$ corresponds to the four neighbouring surface zones.

Furthermore, since the absolute brightness level is unconstrained in the purely spectroscopic Doppler imaging (DI) brightness inversion, the brightness map was biased to $T_0=1$, and values higher than 1 were more strongly suppressed compared to brightness values lower than 1. This was accomplished with the help of the following additional regularisation function
\begin{equation}
R^{(2)}_T = \left\{ \begin{array}{ll}
\sum_{i}\Lambda^{(2)}_T C_1(T_i - T_0)^2, &\mbox{ if $T_i \le T_0$} \\  
\sum_{i}\Lambda^{(2)}_T C_2(T_i - T_0)^2, &\mbox{ if $T_i > T_0$} 
       \end{array} \right. 
,\end{equation}
where $\Lambda^{(2)}_T$ is the second brightness regularisation parameter, $C_1$ and $C_2$ are set to 1 and 10, respectively, and $T_0=1$. 

The values of $\Lambda_B$ and $\Lambda_T$ were determined by trial and error, using the ratio between the mean deviation of the observed and calculated LSD profiles and the mean noise level of observations as a guide. A smaller ratio corresponds to a better fit, but, at the same time, this ratio should not be smaller than 1 since that would indicate over-interpretation of observational data. For our final Stokes $I$ fits, this ratio was 1.81 and 1.98 for the two epochs (May 2014 and January 2017, see below) analysed with DI, and for Stokes $V$ the ratio was 1.05 and 1.06. 

\begin{figure}
\centering
\includegraphics[width=\columnwidth]{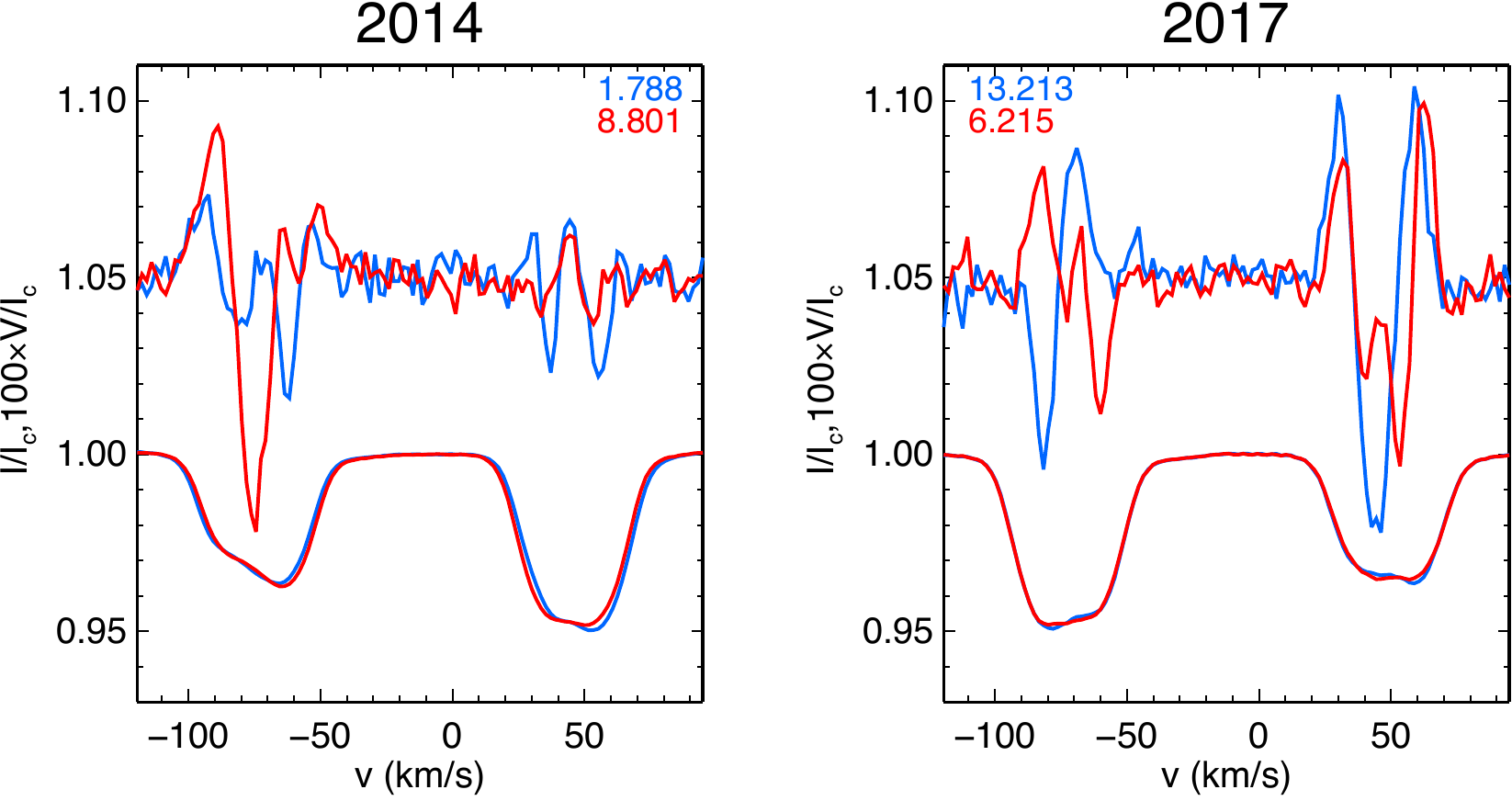}
\caption{Comparison of the LSD Stokes $IV$ profiles from close rotational phases at 2014 and 2017 observing epochs. The bottom profile of each figure shows the LSD Stokes $I$ spectra. The  upper profiles correspond to the LSD Stokes $V$ spectra shifted vertically and magnified by a factor of 100. Each figure shows two observations in red and blue. The respective phases are indicated at the upper right and upper left corner of each figure.}
\label{comp}
\end{figure} 

\begin{figure*}
\centering
\includegraphics[width=0.42\textwidth]{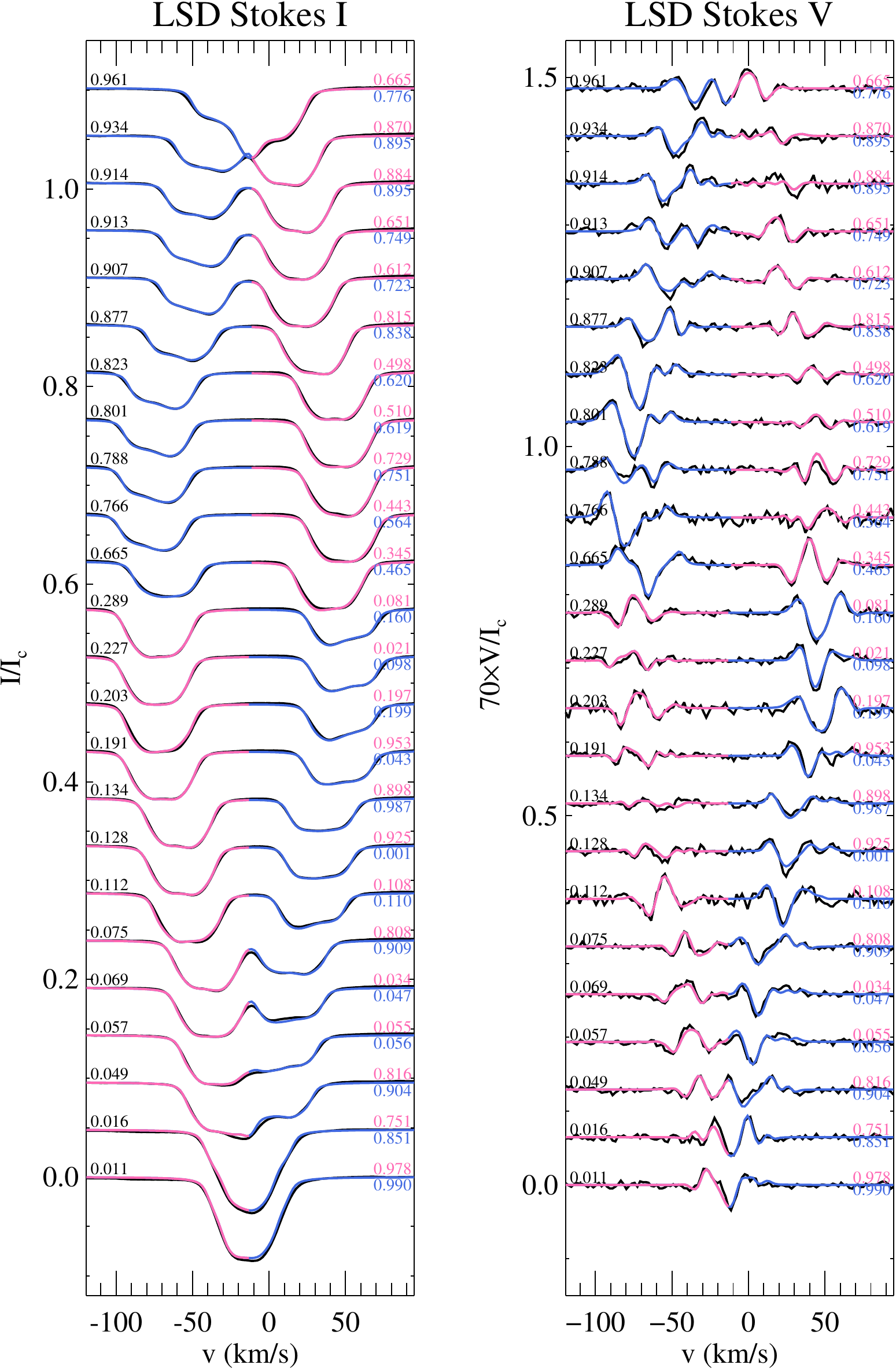}
\includegraphics[width=0.55\textwidth]{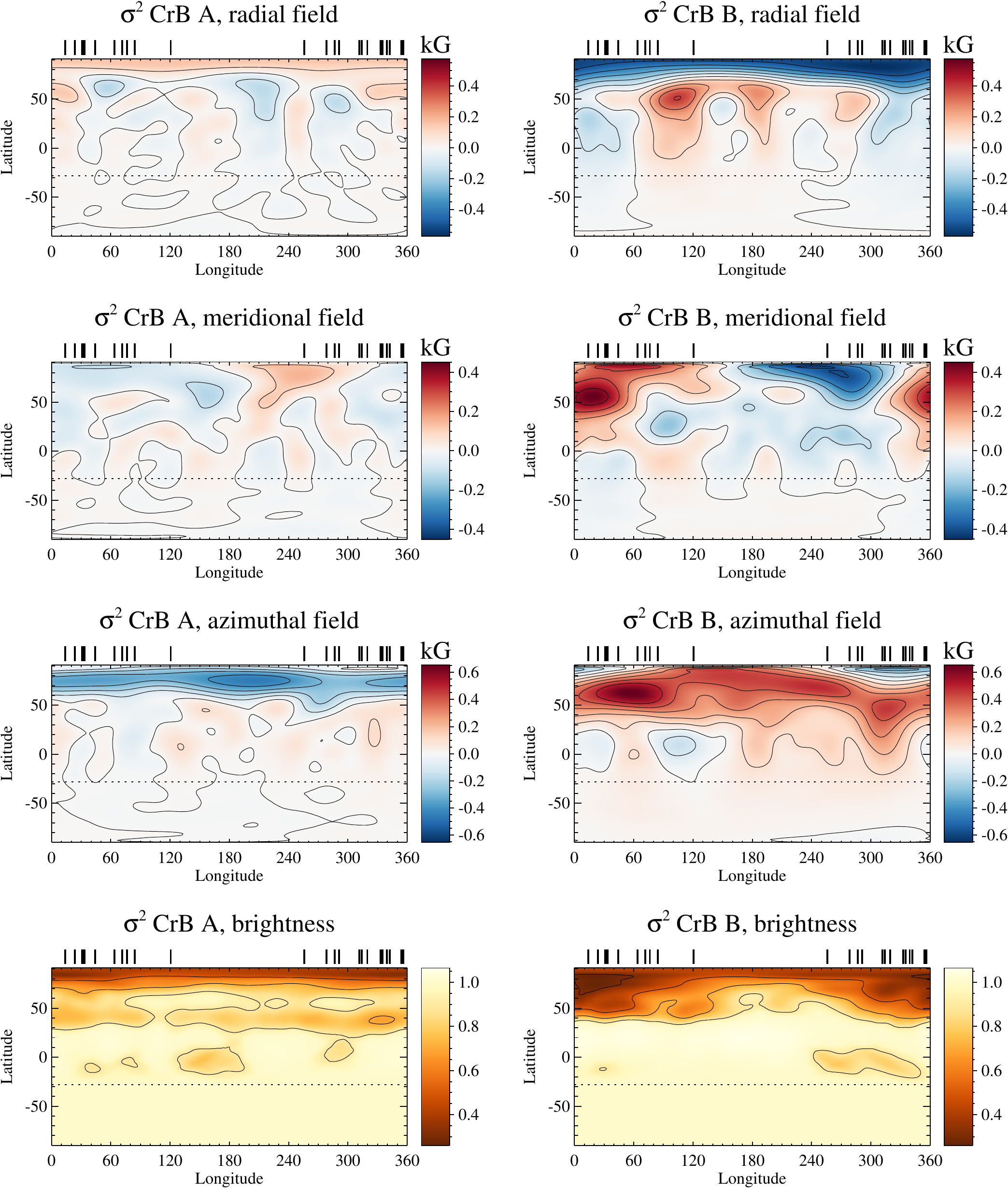}
\caption{LSD Stokes $IV$ line profiles and the reconstructed surface brightness and magnetic maps of $\sigma^2$~CrB components for the 2014 epoch. The black (thin) lines represent the observed LSD profiles, the pink (thick) lines the corresponding model profiles of the primary, and the blue (thick) lines the model profiles of the secondary. The Stokes $V$ profiles were magnified by a factor of 70. The orbital phase is indicated in black to the left above each line profile and the individual rotation phases of the two components are indicated in pink and blue, respectively, to the right of each line profile. The left column of rectangular maps corresponds to the primary star while the right column shows the secondary. The contours in the magnetic maps are plotted with a step of 100~G and in the brightness maps with a step of 0.2 starting at 0.3. The horizontal dotted lines in the rectangular plots show the lowest visible latitude for the inclination angle of $i=28\degr$. The black ticks above each rectangular map indicate orbital phases of observations. Longitude $0\degr$ of the primary is facing longitude $180\degr$ of the secondary.}
\label{linemaps_bm}
\end{figure*} 

\begin{figure*}
\centering
\includegraphics[width=0.42\textwidth]{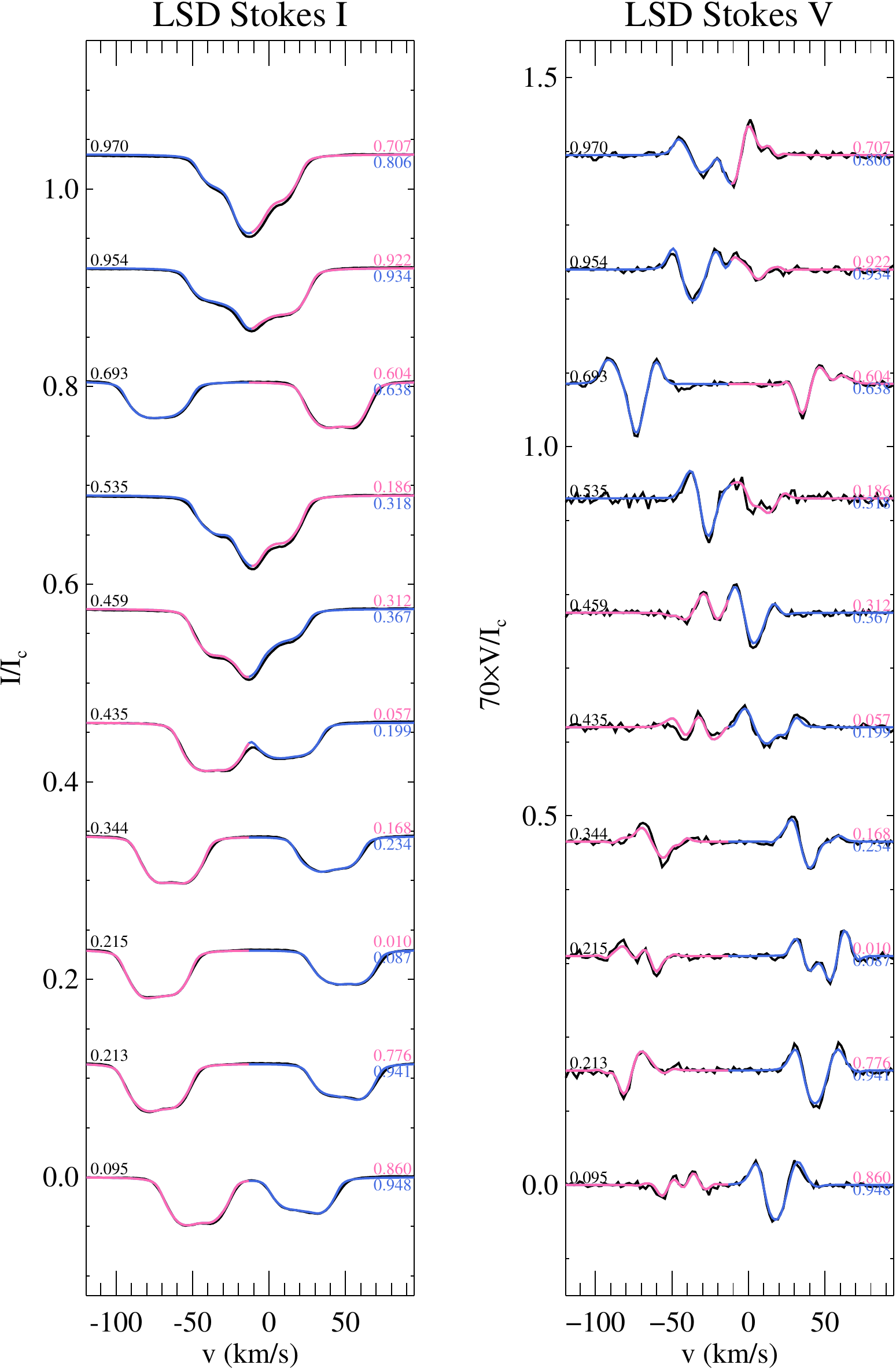}
\includegraphics[width=0.55\textwidth]{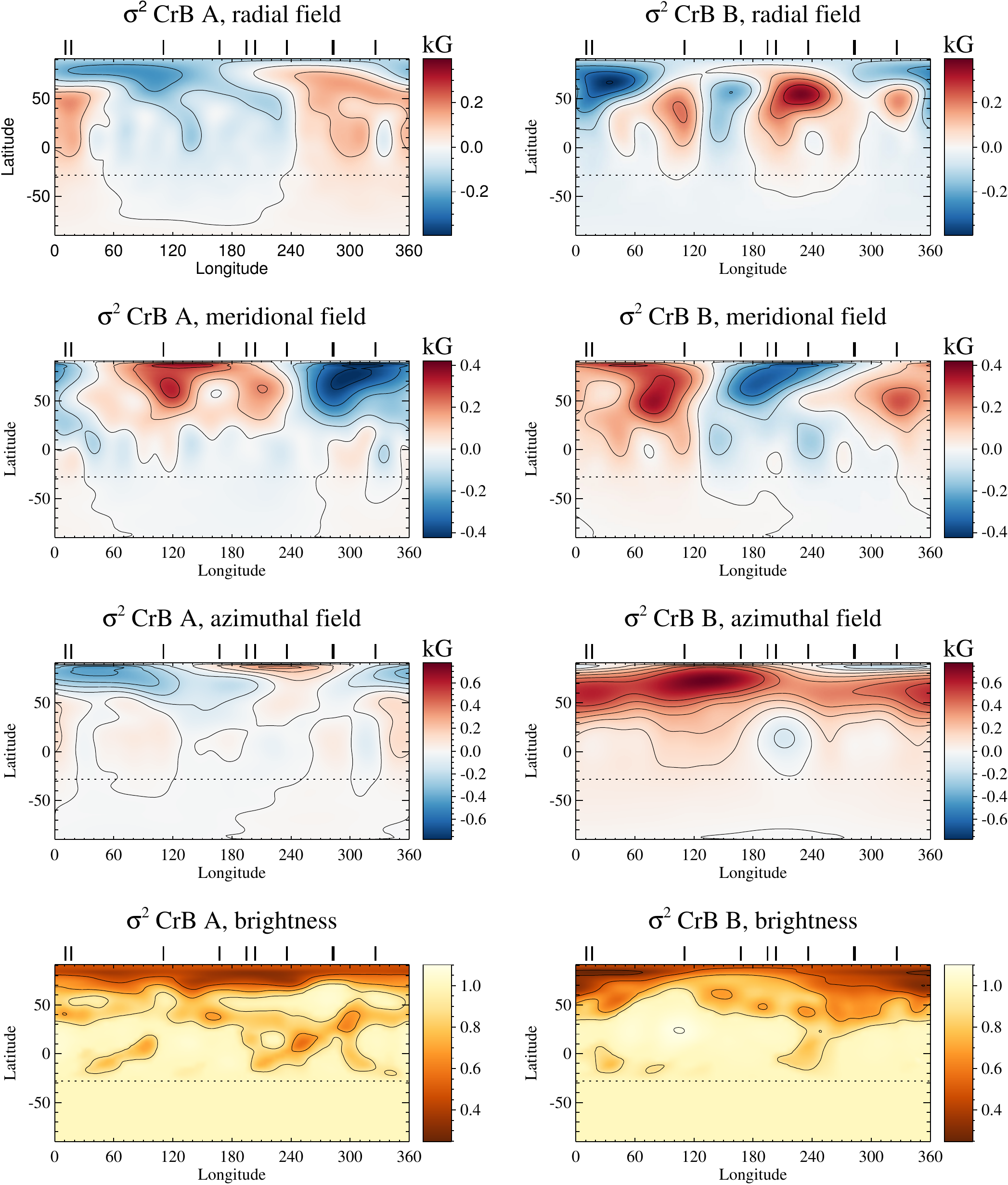}
\caption{Same as in Fig.~\ref{linemaps_bm} but for the 2017 epoch.}
\label{linemaps_mm}
\end{figure*} 

Doppler imaging analysis of temperature spots and magnetic fields on cool active stars is commonly carried out using separate inversions based on the Stokes $I$ and $V$ data, respectively. This kind of analysis was previously applied to the pre-main sequence double-line binary systems V824~Ara \citep{Dunstone08} and V4046~Sgr \citep{Donati2011}. A more accurate approach, implemented in {\sc InversLSDB}, is to perform  magnetic mapping simultaneously and self-consistently with temperature/brightness inversion, i.e. to model the Stokes $I$ and $V$ profiles simultaneously. On the Sun, strong magnetic fields emerging through the surface inhibit convective energy transport, which makes magnetic regions dark and cool compared to the rest of the photosphere. In turn, this affects the intensity spectrum but also the polarisation profiles by reducing their amplitude in the disc-integrated stellar spectrum. If this effect is not taken into account, a small polarisation amplitude might be misinterpreted as a weak magnetic field and the local field strengths are grossly underestimated \citep{Rosen12}. 

Another important shortcoming of most cool star ZDI studies, including the present analysis of $\sigma^2$~CrB, is using the Stokes $V$ data without corresponding linear polarisation profiles. In some situations this leads to a noticeable crosstalk between the radial and meridional field components \citep[e.g.][]{Donati1997,Kochukhov02} and the overall underestimate of the meridional magnetic field strength. Unfortunately, Zeeman linear polarisation has only been detected for a few cool active stars \citep{Kochukhov11,Rosen13} and only one active star has, so far, been successfully mapped using all four Stokes parameters \citep{Rosen2015}.  

In general, it is necessary to analyse multiple observations covering as many rotational phases as possible to map the stellar surface. At the same time, the observations have to be obtained within a time period during which the surface structure does not change significantly. Out of our 43 spectropolarimetric observations of $\sigma^2$~CrB, 3 were obtained during May and June of 2013, another 30 a year later, and the remaining 10 another 2.5 years later. It turns out that LSD profiles, especially Stokes $V$, obtained in the 2014 and 2017 observing epochs exhibit significant differences for the same orbital phases separated by more than 4 orbital periods. This behaviour, illustrated in Fig.~\ref{comp}, indicates either a very fast evolution of the stellar surface structure or a departure of the visible surface rotation from strict synchronicity with the orbital motion.

The 3 observations from 2013 provide a very poor phase coverage. Hence, it was not possible to use these observations for mapping purposes. Of the 30 observations from 2014, 24 were obtained during 11 days in May. The 10 most recent observations from January 2017 span 14 days. We attempted to use all the observations obtained in May 2014 as one ZDI data set and all the observations from January 2017 as another. However, since the observed LSD profiles within each data set could not be satisfactorily phased according to the orbital period, we switched to the second (spherical star shape, non-synchronous rotation) mode of {\sc InversLSDB} and searched for individual rotation periods providing the best fit to the Stokes $V$ observations. Assuming that the rotational axes of both stars are aligned with the orbital axis, we determined $\Delta P=P_{\rm rot} - P_{\rm orb}=0.039\pm0.005$~d ($P_{\rm rot}=1.179$~d) for the primary and $\Delta P=0.024\pm0.005$~d ($P_{\rm rot}=1.164$~d) for the secondary using observations from 2014. This is longer than the orbital period, indicating that rotation of higher latitudes (dominating the visible surface of $\sigma^2$~CrB components given their 28\degr\ inclination) lags behind the orbital co-rotation for both components. This is compatible with a solar-like differential rotation.

To confirm these results we performed inversions using differential rotation as a free parameter. We found that a relatively good fit to the 2014 data can be achieved by adopting equatorial rotation periods equal to $P_{\rm orb}$ and invoking a solar-like differential rotation with $\alpha=\Delta \Omega/\Omega_{\rm e}\approx0.043$--0.045 for both components. On the other hand, our data appears to be insufficient for reliable determination of all four ($\Omega_{\rm e}$ and $\Delta\Omega$ for each of the components) possible differential rotation parameters. For this reason, we present below results of the inversions obtained with $\alpha=0.0$ and individual rotational periods determined above.

\begin{figure*}
\centering
\includegraphics[height=0.8\textwidth,angle=270]{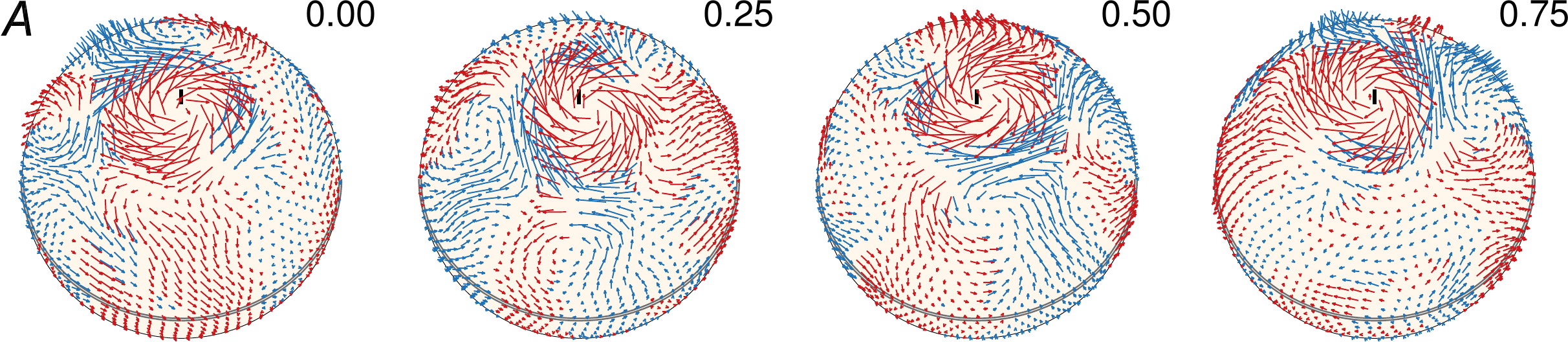}
\includegraphics[height=0.8\textwidth,angle=270]{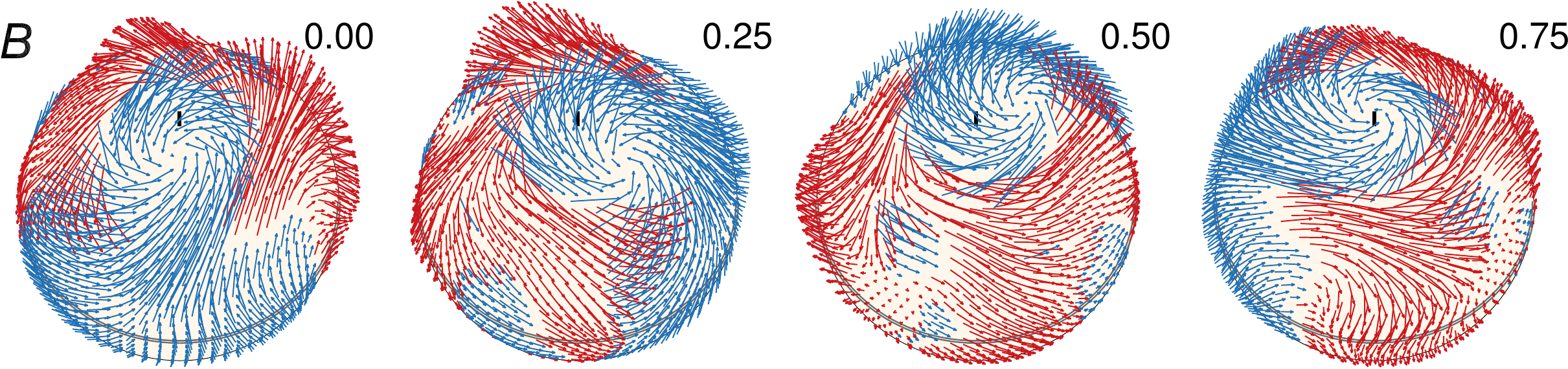}
\caption{Magnetic field vector orientation at the surface of the two components of $\sigma^2$~CrB for the epoch 2014 is shown at 4 rotational phases, indicated above each map. The primary component is in the upper panel denoted $A$, and the secondary component is in the lower panel denoted $B$. The arrow length is proportional to the magnetic field strength. The blue arrows indicate a negative radial magnetic field while the red arrows indicate a positive radial magnetic field. }
\label{vect_bm}
\end{figure*} 

\begin{figure*}
\centering
\includegraphics[height=0.8\textwidth,angle=270]{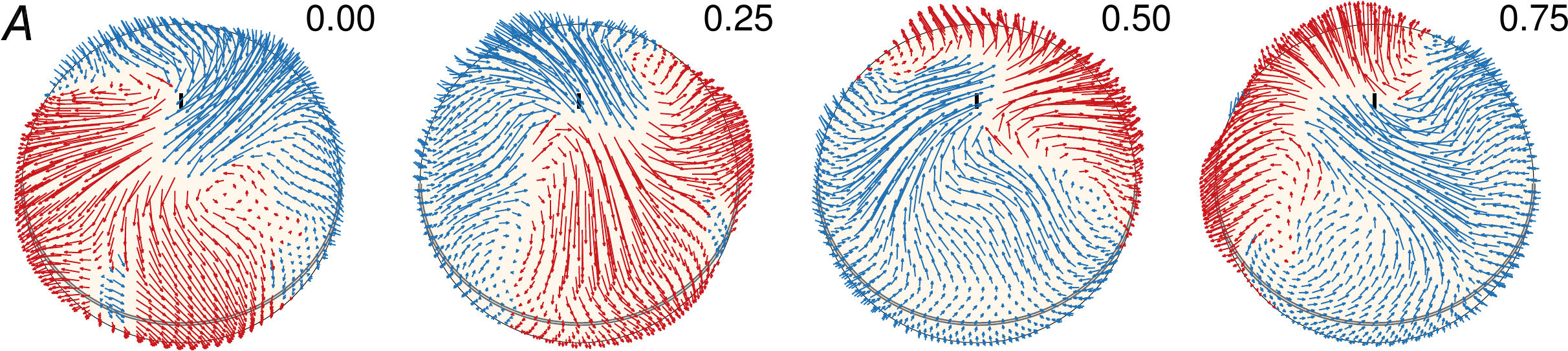}
\includegraphics[height=0.8\textwidth,angle=270]{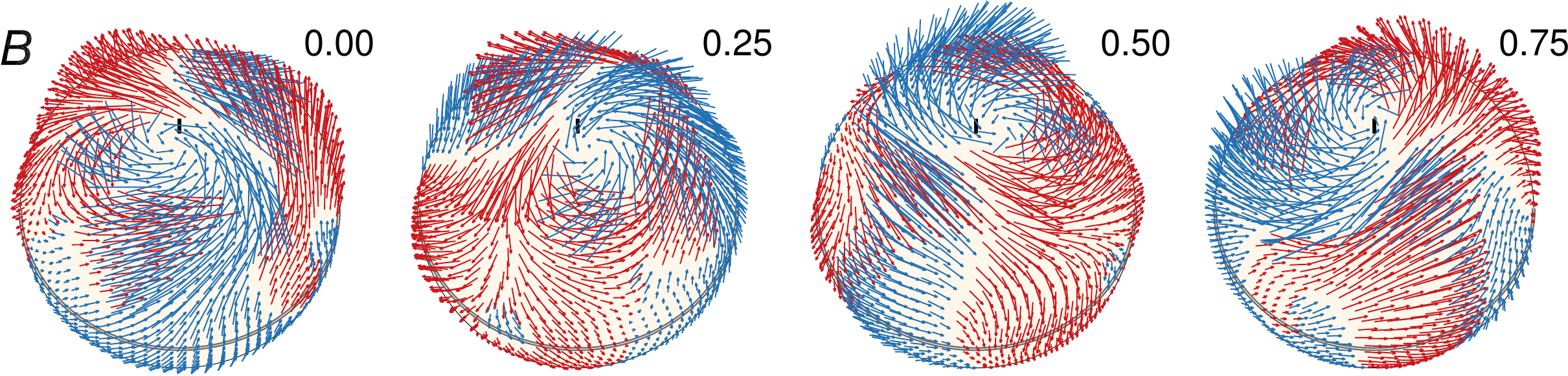}
\caption{Same as in Fig.~\ref{vect_bm} but for the 2017 epoch.}
\label{vect_mm}
\end{figure*} 

\subsection{Results}
\label{res}

Magnetic field and brightness maps were derived for both components of $\sigma^2$~CrB for the observational epochs 2014 and 2017. The resulting line profile fits can be found in Figs.~\ref{linemaps_bm} and \ref{linemaps_mm}. As can be seen from these plots, the Stokes $V$ signatures of the secondary star are systematically stronger than those of the primary. This directly corresponds to a stronger magnetic field of the secondary component, as can be seen in the magnetic maps also presented in Figs.~\ref{linemaps_bm} and \ref{linemaps_mm}. 

Another interesting feature to note in the reconstructed magnetic field maps is the orientation of the magnetic fields at the visible poles of the components. In the 2014 epoch, the primary has a positive radial field at the pole, while the secondary has a negative radial field. The other two field components, meridional and azimuthal, also, approximately, seem to follow the same pattern of opposite polarities. The azimuthal field is dominantly negative for the primary, and dominantly positive for the secondary star. The meridional field configuration at the pole is mixed where close to half is positive and the other half is negative. However, the positive (negative) field of the primary roughly overlaps with the negative (positive) field of the secondary. This can also be seen in Fig.~\ref{vect_bm} where the magnetic field vector orientation is illustrated. However, no such distinctly opposite polarities at high latitudes can be found in magnetic maps for the 2017 epoch.

If the two stars are tidally locked, the same sides always face each other. In the rectangular maps of Figs.~\ref{linemaps_bm} and \ref{linemaps_mm} these sides are centred at longitude $0\degr$ and longitude $180\degr$ for the primary and secondary, respectively, which corresponds to phases 0.0 and 0.50 in the spherical maps in Figs.~\ref{vect_bm} and \ref{vect_mm}. Comparing these opposing sides of the stars, no clear correlations can be found at lower latitudes, approximately below $50\degr$, in either epoch. Since the stars are not visible to us below latitude $-28\degr$, there is no direct information about the field configuration and brightness distribution at these latitudes. Hence, all features below this latitude are highly uncertain, and typically, no strong magnetic features or brightness contrasts are recovered there.

\begin{figure}
\centering
\figps{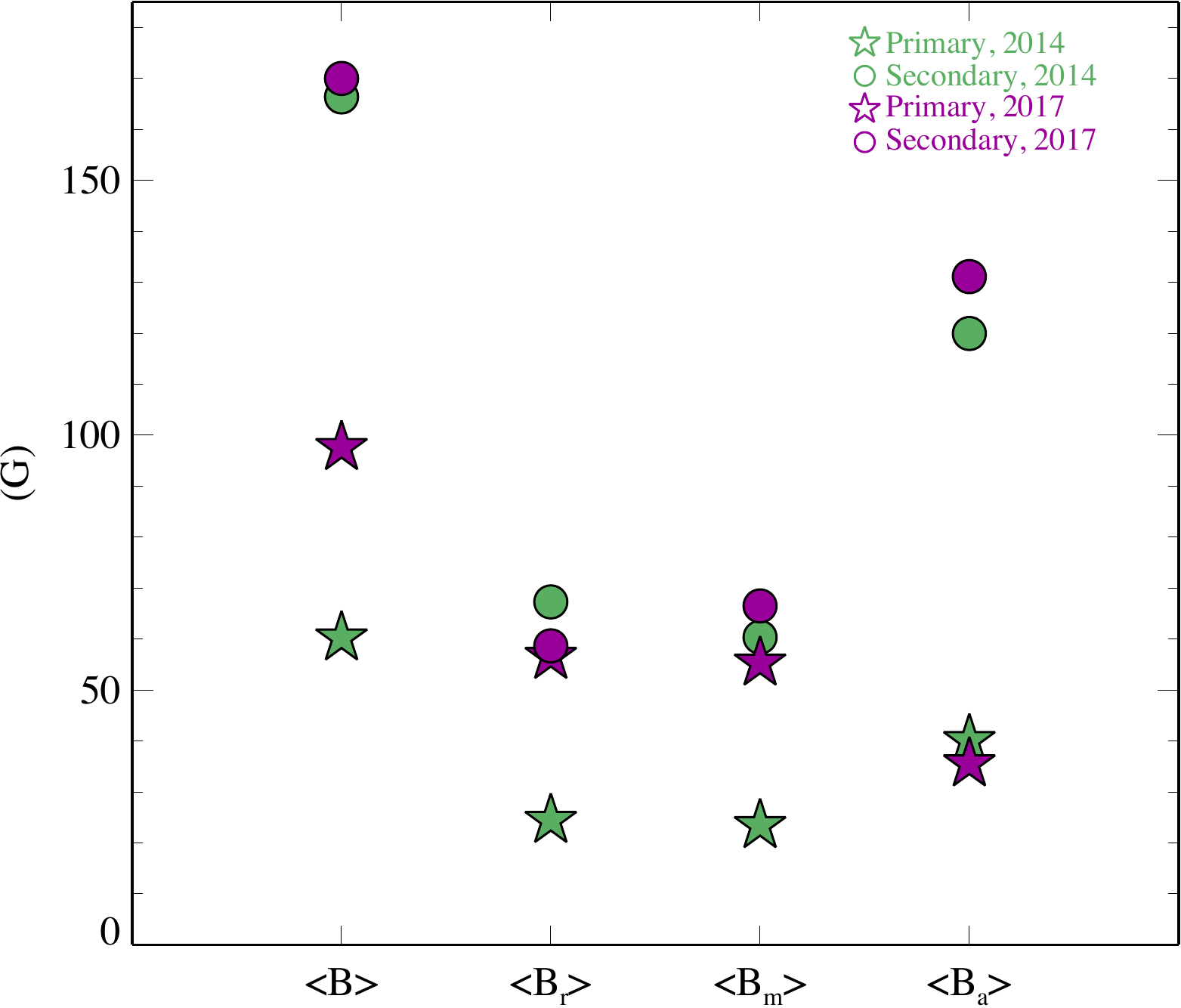}
\caption{Total mean field strength and mean strength of each of the three magnetic field vector components separately. The values for the primary are represented by the stars, while the values for the secondary are represented by the circles. Different colours indicate the 2014 (green) and 2017 (purple) epochs.}
\label{bmod}
\end{figure}

The field strengths of the $\sigma^2$ CrB components can be compared using, for instance, the local maximum absolute field strength, $B_{\rm comp}^{\rm max}$, where $\mathrm{comp}=\mathrm{r}, \mathrm{m}, \mathrm{a}$ for the radial, meridional, and azimuthal magnetic vector component. The single strongest local feature is found for the secondary in both maps. Another way of quantifying the difference between the two stars is by calculating the total mean field strength, $\langle B \rangle= \sum_i S^i \cdot \sqrt{(B_{\rm r}^i)^2+(B_{\rm m}^i)^2+(B_{\rm a}^i)^2}$, and the mean field strength of each magnetic component separately, $\langle B_{\rm comp} \rangle= \sum_i S^i \cdot |B_{\rm comp}^i|$ , where $S^i$ represents the normalised area of surface element $i$. These values are illustrated in Fig.~\ref{bmod}, which shows that the secondary star has a higher field strength, up to 3.7 times, compared to the primary for all components. The total magnetic field energy is 5.7 and 3.3 times higher for the secondary for the 2014 and 2017 epochs, respectively.

\begin{table}[t!]
\caption{Fractions of the toroidal and axisymmetric magnetic field energies.}
\label{menergy}
\centering
\begin{tabular}{ccccc}
\hline\hline
Star &Obs. & $E_{\rm tor}/E_{\rm tot}$  & $E_{m =0}/E_{\rm tot} $ & $E_{m < \ell /2}/E_{\rm tot}$  \\
       & epoch & (\%) & (\%) & (\%) \\
\hline
Primary & 2014 &  72 &61& 74 \\
Secondary & 2014 & 71 &59 &69 \\
Primary & 2017   & 55 &11 & 34 \\
Secondary & 2017 & 74 & 63 & 70 \\                      
\hline
\end{tabular}
\end{table}

We can also compare magnetic maps of the same stellar component for the two epochs. The maximum (unsigned) local field strengths, found for the azimuthal field component, of the primary and secondary stars are $B_{\rm a}^{\rm max}= 459$~G and $B_{\rm a}^{\rm max}= 654$~G, respectively, in the 2014 maps compared to the peak strengths of $B_{\rm m}^{\rm max} = 422$~G and $B_{\rm a}^{\rm max}= 778$~G in the 2017 maps. In Fig.~\ref{bmod} it can be seen that the mean field strength is systematically stronger for the more recent epoch. For the primary star, the difference in field strength is very significant, with the radial and meridional field components up to 2.5 stronger in the later epoch, while the magnetic field strength of the secondary is relatively similar between the two epochs. This also becomes evident if the total magnetic energy is compared. For the primary and secondary, the total magnetic energy is 1.9 and 1.1 times as high in the 2017 epoch, respectively. 

It is also useful to compare the three magnetic field components for the same star to each other. The component $B_{\rm m}^{\rm max}$ is weakest in three out of the four cases and $\langle B_{\rm{m}} \rangle$ is the weakest in the 2014 epoch for both stars. At the same time, $B_{\rm a}^{\rm max}$ is strongest in three out of four cases and $\langle B_{\rm{a}} \rangle$ is strongest in three out of four sets of maps.

\begin{figure}
\centering
\includegraphics[width=0.43\textwidth]{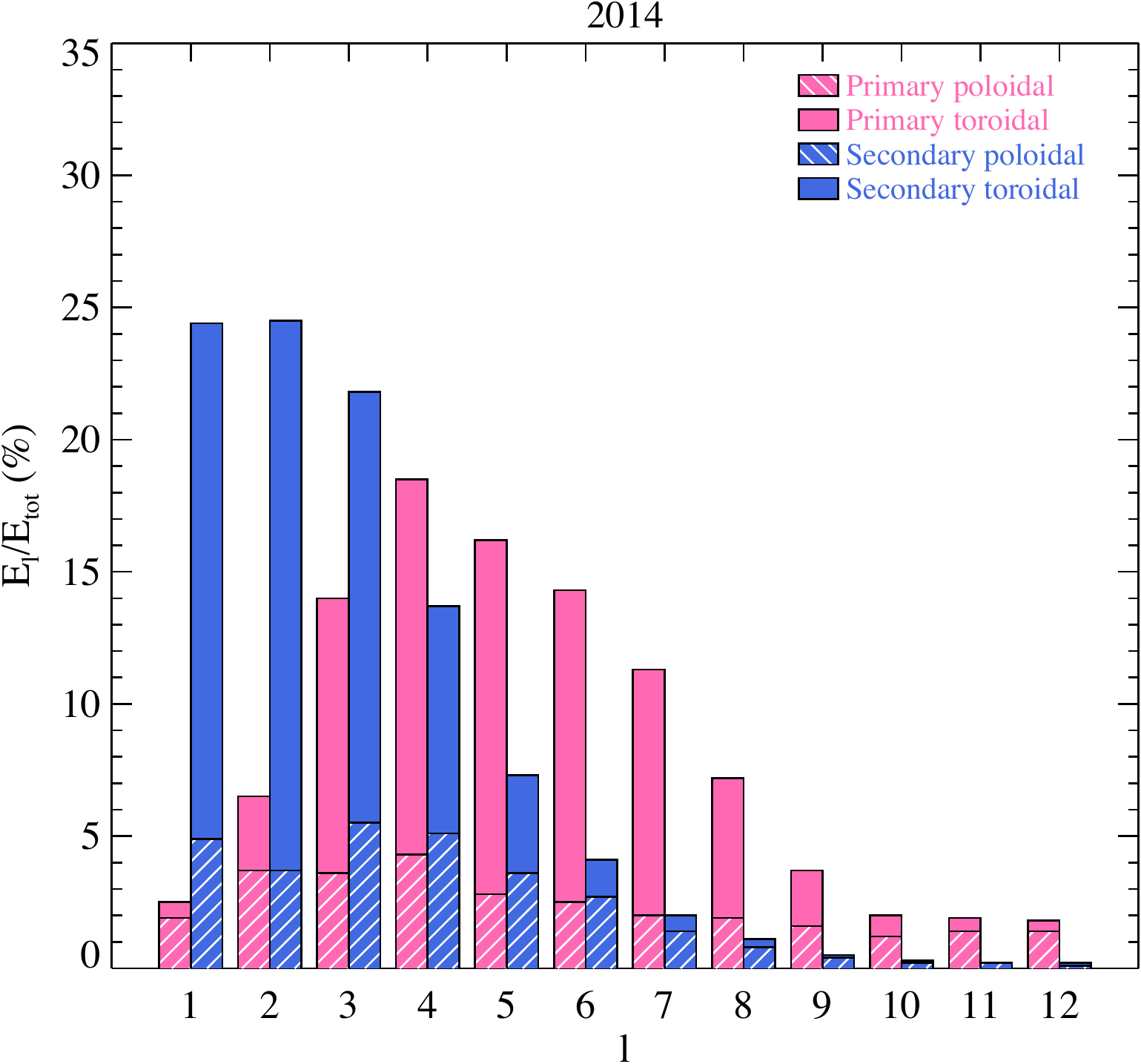}\vspace*{0.3cm}
\includegraphics[width=0.43\textwidth]{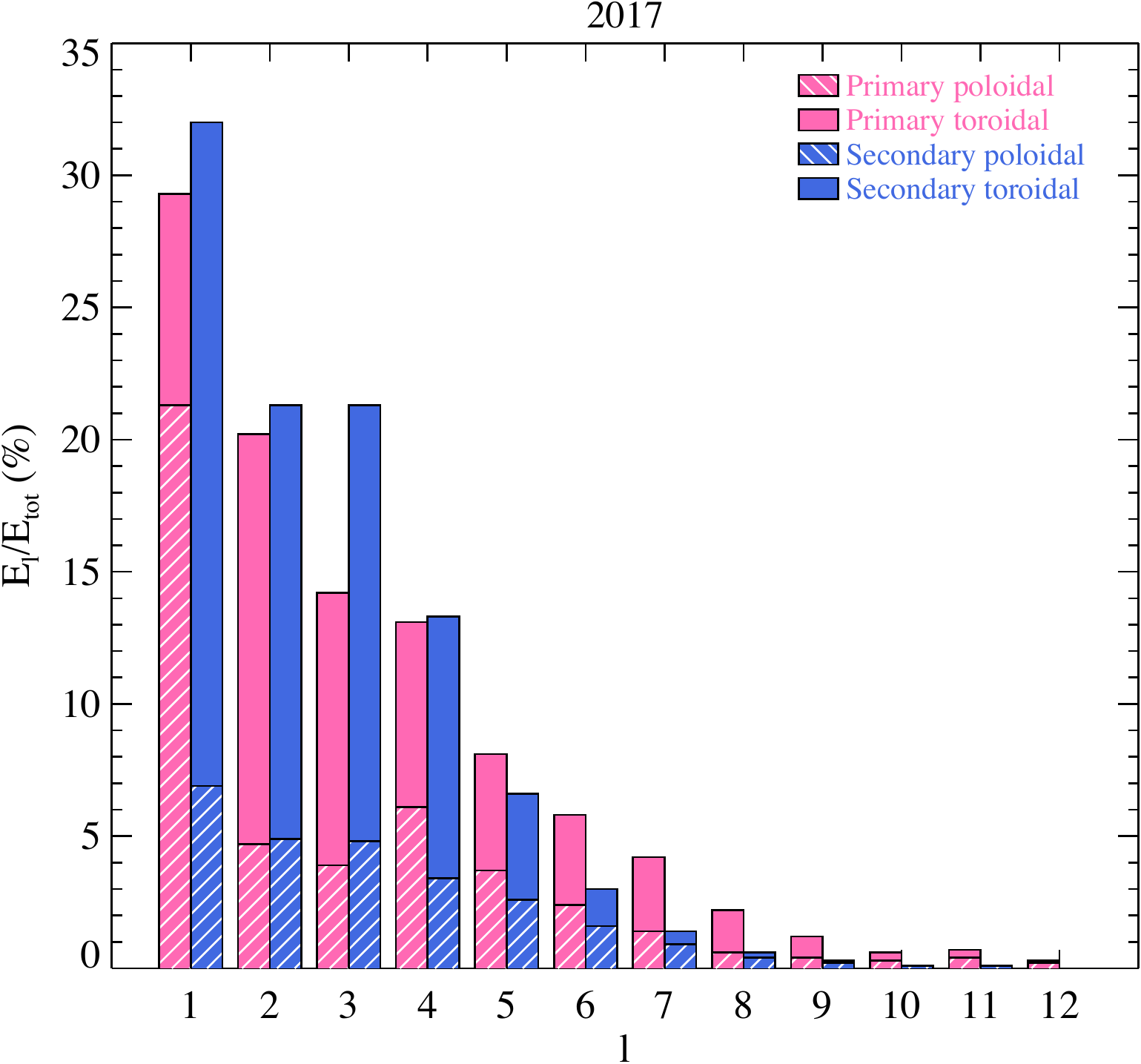}
\caption{Magnetic energy of the poloidal and toroidal field components as a function of the spherical harmonic angular degree $\ell$ for the 2014 (upper panel) and 2017 (lower panel) epochs.}
\label{lmode}
\end{figure}

The distribution of the magnetic field energy over various harmonic modes can be studied in to characterise the complexity of the field topology and assess its evolution. The amount of magnetic energy in each $\ell$ as a fraction of the total magnetic energy, $E_{\rm \ell}/E_{\rm tot}$, can be found in Fig~\ref{lmode}. In the 2014 epoch for the primary, the largest $E_{\rm \ell}/E_{\rm tot}$ is found for $\ell=4$. The dipole component has a lower energy than any $\ell$s between 2 and 9. For the secondary in the same epoch, the dipole and quadrupole components have almost identical energies, whereafter the energy is decreasing with increasing $\ell$. In the 2017 epoch, the dipole component of the primary instead has the highest energy content, similar to that of the secondary for the same epoch. For both stars the energy is systematically decreasing with increasing $\ell$.

The spherical harmonic representation of the magnetic field describes the poloidal, $E_{\rm pol}/E_{\rm tot}$, and toroidal, $E_{\rm tor}/E_{\rm tot}$, components of the field. Both stars show predominantly toroidal magnetic fields in both epochs, even though the primary is less so in the later epoch. The $E_{\rm tor}/E_{\rm tot}$ ratios are listed in Table~\ref{menergy}. 

Moreover, it is possible to assess the axisymmetry of the field. There are two commonly used definitions of axisymmetry. One defines all components with $m < \ell /2$ as axisymmetric, $E_{m < \ell /2}/E_{\rm tot}$ \citep[e.g.][]{Fares2009}. The other is more strict and only assigns components that are exactly aligned with the rotation axis as axisymmetric, i.e. when $m=0$, $E_{m =0}/E_{\rm tot}$ \citep[e.g.][]{See2015}. The primary star is dominantly axisymmetric according to both definitions in the 2014 epoch. In the 2017 epoch, the field topology is changed and the magnetic field is now dominantly non-axisymmetric using both definitions. The secondary star is continuously axisymmetric. The precise values of axisymmetric energy fractions can be found in Table~\ref{menergy}. Both stars seem to deviate from the suggested cool star trend, where $E_{\rm tor}/E_{\rm tot}$ is approximately equal to or less than $E_{m =0}/E_{\rm tot}$ \citep{See2015}. Here, instead, the opposite is seen for the more strict definition in all cases, and the largest discrepancy is for the primary star in the 2017 epoch, where $E_{\rm tor}/E_{\rm tot}=53\%$ and $E_{m =0}/E_{\rm tot}$ is only 11\% and even $E_{m < \ell /2}/E_{\rm tot}$ is only 34\%.

Finally, all Stokes $I$ profiles show clear distortions due to cool spots, and many of these profiles appear to have a flat bottom. This could be an indication that the brightness of the visible pole of each star is lower compared to the rest of the surface. This is confirmed by the reconstructed brightness maps presented in Figs.~\ref{linemaps_bm} and \ref{linemaps_mm}. There are some differences between the two epochs in the structure of surface inhomogeneities, but they show similar brightness contrasts. 

\section{Discussion}
\label{dis}

In previous studies of $\sigma^2$~CrB it was shown that both components have cool temperature spots \citep{Strassmeier2003}. The surface magnetic field was directly detected for the secondary component only \citep{Donati1992} and no quantitative information on the field strength and topology was available for either star. In this study we collected high-quality, time-resolved polarisation observations of this system and were able to detect a magnetic field in both components and then reconstruct the surface brightness and magnetic field distributions for both stars using a newly developed binary-star ZDI code. We found that the magnetic field of the secondary is noticeably stronger compared to that of the primary, especially in the 2014 epoch, as suggested by its higher amplitude circular polarisation profiles. The primary also appears to exhibit a more significant change of its field topology between the two observational epochs separated by 2.3 years.

The difference between the global magnetic fields of $\sigma^2$~CrB A and B can potentially be explained by their long-term activity cycles being out of phase, meaning that there might be periods when the primary star has a stronger magnetic field. Reports of drastic changes in the field strength and topology for stars with parameters similar to $\sigma^2$~CrB can be found in the literature \citep[e.g.][]{Rosen2016}. However, a stronger magnetic field was apparently observed in the secondary already by \citet{Donati1992} based on the data collected in 1990--1991, indicating that the same difference in relative magnetic activity has persisted over nearly 30 years. Also, during all our observations in 2013, 2014, and 2017, the amplitude of the Stokes $V$ signatures of the secondary star is systematically larger compared to that of the primary. All this suggests that magnetic cycles are an unlikely explanation of the different observed magnetic field properties.

A key ingredient for a magnetic dynamo, which is believed to be the main driving mechanism of a cool star magnetic field, is an interplay between convection and rotation. Therefore, it is commonly assumed that stars of similar internal structure, rotation rate, and age must exhibit the same magnetic activity. The two components of $\sigma^2$~CrB have nearly identical mass and radius, the same age and metallicity, and should, therefore, have very similar interior structures. They also have the same, or at least very similar, rotation periods enforced by the orbital motion in a close orbit. Therefore, finding this rather significant and persistent difference in the global magnetic fields of two otherwise very similar stars is surprising. 

This situation is somewhat reminiscent of the recent finding of a major difference in the magnetic field structure for components of an M-dwarf binary containing very similar, low-mass, fully-convective stars \citep{Kochukhov2017}. Similar discrepancies of magnetic field properties, not correlated with any measurable stellar parameters, was previously observed for a sample of single, fully-convective M-dwarf stars \citep{Morin2010}. It was suggested that these differences are linked to the two different regimes of convective dynamo operating at low Rossby numbers, where the stars can have either a strong, axisymmetric, dominantly dipole-like field or a weaker, more complex magnetic field geometry \citep{Gastine2013}.

Our ZDI results show that the secondary component of $\sigma^2$~CrB has a stronger magnetic field compared to the primary. The secondary also has a strong dipole component, with 24--34\% of the energy deposited there, and a dominantly axisymmetric field in both epochs. The primary, on the other hand, has a very weak dipole component in the 2014 epoch, only 3\% of the total magnetic energy. At the same time, its magnetic field is more axisymmetric compared to the secondary. In the 2017 epoch, it is almost the other way around. The dipole component is now the strongest, comparable to the dipole component of the secondary, but the field is instead predominantly non-axisymmetric. This is a more complex picture compared to what has been observed in fully convective stars and, at this stage, it is somewhat speculative to attribute this behaviour to the similar bi-stable dynamo phenomenon operating at a higher stellar mass. Nevertheless, our ZDI results demonstrate beyond any doubt that the mechanism generating magnetic field in the interiors of rapidly rotating Sun-like stars, such as $\sigma^2$~CrB components, is not a straightforward function of fundamental stellar parameters and rotation rate.

It is interesting to note that the radial magnetic fields on the visible poles of $\sigma^2$~CrB components have opposite polarity in the 2014 epoch, although not in the 2017 epoch. In the earlier epoch, the opposite polarity is also, approximately, seen for the azimuthal and meridional field components. This could suggest that the magnetic field lines extending from the poles are connecting the two stars. The potential source surface field (PSSF) extrapolation technique can be used to derive the magnetospheric structure of a star. \citet{Holzwarth2015} developed an extended PSSF model suitable for analysis of interacting magnetospheres of close binary stars. They  used this technique for V4046~Sgr \citep{Gregory2014,Holzwarth2015}, which is a PMS binary comprising two classical T Tauri stars in a circular, synchronised 2.4~d orbit. Their calculations were based on the radial ZDI maps reconstructed by \citet{Donati2011}. The magnetic field strengths of these two stars are comparable to those found for $\sigma^2$~CrB, i.e. a few hundred G. The magnetic topologies of V4046~Sgr components are predominantly dipolar, but tilted with respect to the rotation axes, which are aligned with the orbital axis. \citet{Holzwarth2015} found that components are connected by the field lines emerging from the lower latitudes at the inner part of the binary. The stellar rotation and the orbital revolution of the $\sigma^2$~CrB system also appear to be approximately synchronised, and the stellar parameters are compatible with the notion that the rotational axes are aligned with the orbital axis. The field geometry of $\sigma^2$~CrB in 2014 suggests the existence of a magnetospheric connection between the poles instead of at lower latitudes. This connection does not seem to be continuous, however, since the 2017 maps do not show any clear evidence of opposite magnetic field polarity between the two stars.

The spectra for the two observational epochs analysed in detail were obtained during 11 and 14 days, or approximately during 9--12 orbital cycles. Line profiles corresponding to similar orbital phases separated by more than about 4 cycles are clearly discrepant for both epochs. This discrepancy could either be due to a rapid intrinsic evolution of the global magnetic field or  a differential rotation. We showed that the entire sets of spectra obtained in 2014 and 2017 can be successfully modelled by adopting rotational periods that are $\sim$\,0.02--0.04~d longer than the orbital period. The equatorial rotation of the $\sigma^2$~CrB components is expected to be synchronised with the orbital motion since theoretical studies predict synchronisation to occur before the orbital circularisation \citep{Zahn77,Tassoul92}. Consequently, we consider a solar-like differential rotation to be the most likely interpretation of longer apparent rotation periods required to phase our data. Earlier \citet{Strassmeier2003} found that the photometric period of the unresolved $\sigma^2$~CrB system was 0.0172~d longer than the orbital period. They argued that this difference is not due to asynchronism but instead is a sign of differential rotation. This conclusion qualitatively agrees with our results. The minimum shear of $\Delta\Omega$\,=\,0.10--0.19 rad\,d$^{-1}$ corresponding to the difference between the orbital and inferred near-polar rotation periods of $\sigma^2$~CrB components approximately agrees with the empirical differential rotation versus $T_{\rm eff}$ trends discussed by \citet{Barnes05} and \citet{Reiners06}, although one might expect close binary components to follow a dependence different from that of single stars.

Magnetic maps of the $\sigma^2$~CrB components exhibit systematically lower field strength in the meridional field component. This feature is commonly seen in ZDI studies of other cool active stars, which do not include linear polarisation. Numerical tests suggest that ZDI based on the Stokes $V$ data alone tends to underestimate meridional field. Therefore, we cannot ascertain whether the weakness of the meridional field is linked to the dynamo operation or represents a bias of the magnetic field reconstruction procedure.

Global magnetic fields of cool active stars studied with ZDI inversions appear to follow a trend in which $E_{\rm tor}/E_{\rm tot} \lesssim E_{\rm axisym}/E_{\rm tot}$ \citep{See2015}. We estimated the axisymmetry of the field using two different definitions. The results show that both components of the $\sigma^2$~CrB system seem to deviate from this trend, especially the primary star in the 2017 epoch. This suggests that the field topologies of close binary stars with components of similar mass might follow different patterns compared to single active stars.

This is the first ZDI study of a cool system by the BinaMIcS collaboration. We plan to extend this analysis to other interesting systems to compare with $\sigma^2$~CrB and further our understanding of the dynamo processes in binary systems.

\begin{acknowledgements}
O.K. acknowledges financial support from the Swedish Research Council and the Swedish National Space Board.  O.K. is also supported by the project grant ``The New Milky Way'' from the Knut and Alice Wallenberg foundation. 
E.A., C.N., and J.M. acknowledge financial support from ``le Programme National de Physique Stellaire'' (PNPS) of CNRS/INSU, France. E.A. acknowledges financial support from ``le Laboratoire d'Excellence OSUG@2020 (ANR10 LABX56)'' funded by ``le programme d'Investissements d'Avenir''.
G.A.W. acknowledges Discovery Grant support from the Natural Sciences and Engineering Research
Council (NSERC) of Canada.
We thank the CFHT staff for their help and commitment throughout the observing runs.
\end{acknowledgements}


\begin{appendix}
\section{Global continuum normalisation}
\label{contin}

In this paper we carried out continuum normalisation of the entire stellar spectrum using an iterative algorithm based on the optimal filter function. The latter is determined by solving the minimisation problem constrained by the Tikhonov regularisation functional
\begin{equation}
\Phi = \sum_{i=1}^N w_i (y_i - f_i)^2 + \Lambda \sum_{i=1}^{N-1} (f_{i+1} - f_{i})^2 \to \mathrm{min},
\label{eqn:opt}
\end{equation}
where $y_i$ is the observed spectrum, $f_i$ is the fitted curve, $w_i$ are weights assigned to each pixel, and $\Lambda$ is the Tikhonov regularisation parameter. The weights $w_i$ take into account error bars of the observed spectrum and allow one to exclude certain regions (e.g. containing broad absorption lines or deep telluric features) by setting $w_i=0$. The function $f$ is determined from Eq.~(\ref{eqn:opt}) by solving the system of sparse, band-diagonal linear equations
\begin{equation}
\dfrac{\partial \Phi}{\partial f_i} = 0.
\end{equation}

The continuum normalisation procedure starts by merging individual echelle orders into continuous spectrum and interpolating it onto an equidistant wavelength grid. Then the optimal filter function is fitted to the entire spectrum and an asymmetric sigma-clipping is applied to reject many more spectral points below the fitted curve than above it. This has an effect of pushing the fit to higher spectral points. The optimal filter fit followed by the asymmetric spectral pixel rejection is repeated a given number of times or until the standard deviation of the fit reduces to a prescribed value. The final continuum fit is then interpolated back onto the original wavelength grid of the observed spectrum and is applied to individual echelle orders.

The procedure described above does not require manual identification of continuum regions. This algorithm is very effective in automatically isolating and fitting the upper envelope of an arbitrary observation with only two free parameters, the number of iterations, and the smoothing factor $\Lambda$.

\end{appendix}

\end{document}